%% file: 00_main_for_arxiv_submission.tex
\documentclass[preprint,journal]{vgtc}       

\ifpdf
  \pdfoutput=1\relax                   
  \usepackage{graphicx}                
  \DeclareGraphicsExtensions{.pdf,.png,.jpg,.jpeg} 
\else
  \usepackage{graphicx}                
  \DeclareGraphicsExtensions{.eps}     
\fi%

\graphicspath{{figures/}{pictures/}{images/}{./}} 

\usepackage{microtype}                 
\PassOptionsToPackage{warn}{textcomp}
\usepackage{textcomp}
\usepackage{mathptmx}
\usepackage{times}

\usepackage{cite}
\usepackage{tabu}
\usepackage{booktabs}

\usepackage{hyperref}
\usepackage{url}
\usepackage{xcolor}

\usepackage{graphicx}
\usepackage{epstopdf}
\usepackage{booktabs}
\usepackage{float}
\usepackage{caption}

\usepackage{amsmath}
\usepackage{bm}
\usepackage{algorithm}
\usepackage{algpseudocode}
\usepackage{amsmath}
\usepackage{multirow}
\usepackage{indentfirst}
\usepackage{subfig}
\usepackage{epsfig}
\usepackage{epstopdf}
\usepackage{xspace}

\usepackage{subeqnarray}

\usepackage{color}
\usepackage[utf8]{inputenc}
\usepackage{siunitx}
\usepackage{rotating}

\usepackage{paralist}
\usepackage{cleveref}

\usepackage[font={scriptsize,sf}]{caption}
\usepackage{multicol}
\title{A Visual Analytics System for Multi-model Comparison on \\ Clinical Data Predictions}

\author{Yiran Li\thanks{Corresponding author. E-mail: ranli@ucdavis.edu}~\thanks{Department of Computer Science, University of California, Davis}~, Takanori Fujiwara\footnotemark[2]~, Yong K. Choi\thanks{Betty Irene Moore School of Nursing, University of California, Davis}~, Katherine K. Kim\footnotemark[3]~\thanks{School of Medicine, University of California, Davis}~, and Kwan-Liu Ma\footnotemark[2]}

\input{for_arxiv_submission/0_abstract.tex}

\keywords{Clinical data, XAI, tree-based machine learning models, model consistency, measures of dependence, visual analytics}

\vgtcinsertpkg

\begin{document}

\firstsection{Introduction}
\maketitle

\input{for_arxiv_submission/1_introduction.tex}
\input{for_arxiv_submission/2_related_work.tex}
\input{for_arxiv_submission/3_research_questions.tex}

\input{for_arxiv_submission/4_methods.tex}
\input{for_arxiv_submission/5_interface.tex}
\input{for_arxiv_submission/6_case_studies.tex}
\input{for_arxiv_submission/7_discussion.tex}
\input{for_arxiv_submission/8_conclusions.tex}

\section*{Acknowledgments}
This research is sponsored in part by the U.S. National Science Foundation through grant IIS-1741536 and
a 2019 Seed Fund Award from CITRIS and the Banatao Institute at the University of California.

\bibliographystyle{abbrv-doi}
\bibliography{00_main.bib}

\end{document}

%% file: for_arxiv_submission/0_abstract.tex
\abstract{
There is a growing trend of applying machine learning methods to medical datasets in order to predict patients' future status. Although some of these methods achieve high performance, challenges still exist in comparing and evaluating different models through their interpretable information. Such analytics can help clinicians improve evidence-based medical decision making. In this work, we develop a visual analytics system that compares multiple models' prediction criteria and evaluates their consistency. 
With our system, users can generate knowledge on different models' inner criteria and how confidently we can rely on each model's prediction for a certain patient.
Through a case study of a publicly available clinical dataset, we demonstrate the effectiveness of our visual analytics system to assist clinicians and researchers in comparing and quantitatively evaluating different machine learning methods.
}

%% file: for_arxiv_submission/1_introduction.tex
\label{sec:introduction}

Comprehensive health data plays an important role in the provision of high-quality medical care and decision making. 
For example, predictive analysis on medical datasets can help clinicians understand the potential risks of an operation for patients with particular characteristics~\cite{kawaler2012} or identify potential deterioration of health status over various treatment periods~\cite{mortalitypred}. 
To develop useful predictions, machine learning (ML) methods have been used on medical datasets.
For clinical prediction tasks, ML methods must satisfy high accuracy when clinicians rely on the predicted results for their decision-making process~\cite{CDSS}.  

However, accuracy may be necessary but not sufficient. The black-box nature of ML methods creates a concern when it comes to critical medical decision making.
For instance, Caruana et al.~\cite{caruana2015intelligible} trained a neural network (NN) to identify pneumonia patients who should be admitted to hospitals as opposed to being treated in outpatient clinics based on their risk profile. 
Even though this NN model had a high accuracy rate, the model determined that pneumonia patients with asthma should not be admitted, thinking that these patients have a lower risk of dying.
This dubious prediction is caused by the fact that these severe patients had been aggressively treated in intensive care units and, as a result, they survived at a high rate~\cite{caruana2015intelligible,adadi2018peeking}. 
As shown in this example, the decision relied on an ML model might cause critical harm to patients.
There is little tolerance for errors in medical diagnosis or decision making; thus, providing transparency or interpretability in ML models (e.g., explainable AI (XAI)), is critical in validating and reasoning of the resulting predictions.

To address this problem, data scientists have developed several XAI methods to provide interpretable information of ML models~\cite{adadi2018peeking}. 
For example, these methods can show which features of a patient's record have a strong influence on the decision from an ML model (called local feature contributions~\cite{Palczewska2014}). 
This information is useful for both clinicians and researchers to judge whether they can trust the model's decision, using their medical knowledge. 
However, when they want to select the best ML model from various options~\cite{kawaler2012,zheng2017machine}, checking each model's interpretable information on all patients' records for each ML model is not realistic.
Therefore, clinicians and researchers need 
analysis methods that facilitate a comprehensive comparison of the interpretable results across various ML models and different patients.

Although ML models might have similar performances (e.g., high accuracy) in predictions, the models may vary in terms of their algorithms or the insights learned from the dataset. 
To understand such variation, the first step is to obtain an overview and a comparison of the models' inner prediction rationales. 
Then, to evaluate the reliability of each model, focusing on the consistency of inner criteria, the models need to be further compared to another group of similar patients. 

To address the aforementioned needs, we develop a visual analytics system for multi-model comparison based on the interpretable information. 
We focus on analyzing models of tree-based ML methods (e.g., random forest~\cite{RF} and gradient boosting~\cite{GB}) because 
these models are widely used in predictions of clinical datasets~\cite{tree-based1,tree-based2,tree-based3}.  

We use model-agnostic interpretation methods~\cite{adadi2018peeking} 
for systematic comparisons, in conjunction with several unsupervised methods.
Specifically, across the models, we visualize the similarities of local feature contributions for each patient with a dimensionality reduction method. 
This visualization can provide an overview of how differently the model(s) predicted for each patient's result.
Additionally, our system offers a quantitative method to evaluate each model's consistency by utilizing ``measures of dependence''~\cite{reshef2016measuring} developed for statistics.
By visually comparing the consistency of each model's reasoning, users can obtain knowledge of the models' reliability.
Moreover, the information of consistency can help judge whether the users should rely on a predicted result based on which features or which range of feature values contribute to the model's prediction. 

We demonstrate the usefulness of our system by comparing six models tuned with different tree-based ML methods. Through the demonstration, we discuss the feasibility of applying visual analytics upon interpretable results to aid clinicians' model and feature selection.

%% file: for_arxiv_submission/2_related_work.tex
\section{Related Work}
We survey the relevant works in ML-based prediction on clinical data, interpretation methods for ML models, visualizations for XAI, and consistency measurements of prediction models.

\subsection{Machine-Learning Based Prediction on Clinical Data}
For medical predictions on clinical data, two major categories of ML methods are popularly used: tree and recurrent neural network (RNN) based methods.

Tree-based methods are appropriate for medical prediction tasks since they characterize decision-making after encountering different attribute values for each patient.
Besides the nature of prediction tasks, many clinical datasets are in the form of temporal sequences. 
In this sense, the tree-based methods are used to forecast a result at a certain future time step based on the results in past time steps.
Ever since the decision tree~\cite{DT}, much research has been carried out to enhance this method. 
Random forest~\cite{RF} makes use of an ensemble of trees to eliminate the overfitting problem. 
Gradient boosting~\cite{GB} extends the idea of random forest in a way that tries to boost an ensemble of weak learners. 
Further, a series of variations~\cite{XGB,LGB,CGB} in the implementation of gradient boosting are developed to provide more efficient computations and better performance.

RNN-based methods are designed for temporal classification and can thus be used to make predictions on clinical data. 
For example, Lipton et al.~\cite{LSTM} used an RNN with long short-term memory to classify diagnoses of patients' temporal sequences. 
Che et al.~\cite{missing} further enhanced RNN to address a problem of data with missing values.
While deep learning methods are able to provide high prediction performance, these methods typically need a large dataset (e.g., more than tens of thousand records) for their training.

\subsection{Interpretation Methods for Tree and Deep-learning based Models}
Various interpretable ML and post-hoc analysis methods have been developed. 
Adadi and Berrada~\cite{adadi2018peeking} provided a comprehensive survey of interpretation methods. 
Here, we describe only the methods related to tree-based or deep-learning based models. 

For tree-based models, interpretation methods can be categorized as a \textbf{global} scale, which tries to investigate each feature's impact on all instances' predictions (global feature contributions), and a \textbf{local} or \textbf{instance} scale, which describes the contribution of each feature for each instance's prediction (local feature contributions). 
There are several ways to calculate feature contributions. 
For example, Palczewska et al.~\cite{Palczewska2014} introduced a method calculating the local increment value from a parent to a child node corresponding to one feature. Another example is the TreeInterpreter~\cite{treeinterpreter}, which obtains each feature's contributions by going back through each decision path from a leaf to its root.

As for deep learning used for medical predictions, Choi et al.~\cite{Retain} built a variation of RNN, called RETAIN. 
By integrating the attention mechanism in RNN, RETAIN provides interpretable results while keeping prediction performance of RNN.
Also, Kwon et al.~\cite{RetainVis} modified RETAIN so that it also provides the attention value for each feature for each time step.

Besides interpretation methods for specific models, there are also model-agnostic methods.
Model-agnostic methods perform generic model explanations with common components of different models. LIME~\cite{LIME} is an example of an interpreter for any type of model.
Another example is SHAP~\cite{SHAP}, which presents the impact of each feature value on the prediction.
For a more comprehensive list of model-agnostic methods, refer to the survey by Adadi and Berrada~\cite{adadi2018peeking}.

\subsection{Visualizations for XAI}

Liu et al.~\cite{LiuTowards2017} surveyed the recent progress on visualizations developed to understand, diagnose, and refine ML models. 
For example, Wang et al.~\cite{WangDeepVID} developed an interpretation approach to review the inner mechanisms of complicated deep neural networks. 
Manifold~\cite{zhang2018manifold} is a framework for visually interpreting, debugging, and comparing ML models.
RuleMatrix~\cite{MingRuleMatrix} provides rule-based explanations to allow users with little ML knowledge to navigate and validate ML models.
As described in \cite{LiuTowards2017}, while there are many more related works, in the following, we focus on tree- and RNN-based models, which are commonly used for clinical data.

Several works provided visual analytics methods for tree-based models. 
For example, Zhao et al.~\cite{iForestZhao} developed a comprehensive interface for interpretations of random forest. 
Their visual analytics system focuses on model simplification and decision path extraction for a selected group of patients.
Another example is TreePOD~\cite{TreePOD} which is developed for aiding decision tree selection through visually exploring candidate trees. 
Collaris et al.~\cite{instance} performed a case study on instance-level visual reasoning of random forest in the scenario of fraud detection.

As for RNN-based methods, for example, Jin et al.~\cite{CarePre} built a clinical decision assistance system with RETAIN. 
For one selected patient, their system provides which past events have a strong influence on the ML decisions, potential treatment outcomes, and a summary of similar patients' health records. 
These sets of information help clinicians make their decision with confidence.
Guo et al.~\cite{VisProgression} created a scalable interface to aggregate event sequence records of patients based on the RNN model they devised. 
Wang et al.~\cite{FeatureAttr} produced a matrix of small multiples to visually reason about feature attributes (i.e., attention values of their RNN model) as well as a time sequence view to make comparisons. 

Although numerous visual analytics methods have been developed~\cite{LiuTowards2017} , methods for tree-based models still have not been fully studied. 
More specifically, methods for comparing multiple models based on their interpretable information (e.g., local feature contributions) are still missing. 

\subsection{Consistency Measures of Model Interpretations}
On the most generic level, a machine learning algorithm's consistency can be defined as its stability when there are small perturbations of the input data. 
A learning algorithm's stability was first investigated by  Devroye and Wagner~\cite{Devroye1979}, in which they observed the quantitative results of leave-one-out error. 
Since then, many works have been done on probabilistic analysis of learning algorithms' stability~\cite{Devroye1991,Lugosi1994,Kearns1999}. 
Then, Bousquet and Elisseeff~\cite{Bousquet2002} defined the notion of uniform hypothesis stability to derive the generalization error bounds.

In addition to the probabilistic analyses on stability, statisticians also developed the definition and theories of learning algorithms' consistency. 
With the definition by Kearns and Vazirani~\cite{Kearns1994}, an algorithm is consistent if it always returns a result that is consistent with the given examples. 
Then, the definition of PAC (Probably Approximately Correct) learning algorithm~\cite{Haussler1995} is introduced, 
which is an extension of consistent algorithms.

In our visual analytics scenario, the consistency of each ML method's inner rationales is of our interest. 
Therefore, besides the stability or consistency of the learning algorithm, we also survey the measures of dependency between two random variables, which characterize the dependence of an ML model's rationales (e.g., local feature contributions) on the actual feature values.
To begin with, Pearson's correlation coefficient provides a measure of the linear relationship between two variables~\cite{Pearson1895}. 
Later on, Spearman~\cite{Spearman1987} extended Pearson's correlation to nonlinear relationships. 
However, the Spearman's correlation coefficient is limited to monotonic dependencies between variables.
With the development of information theory, 
there are works summarizing the shared information between the two random variables.
For example, mutual information~\cite{Cover2006}, 
maximal information coefficient~\cite{Reshef2011}, 
and total information coefficient~\cite{reshef2016measuring} have been introduced.
All these measures do not have assumptions of the random variables' distributions and, thus, these measures are more suitable when the distributions are unknown and the relationships are nonlinear.

%% file: for_arxiv_submission/3_research_questions.tex
\section{Analysis Questions (AQs)}

As mentioned in the previous sections, the general goal of this study is to \textit{leverage visual analytics when reasoning and comparing multiple models' interpretations}. 
Here, we list more detailed analysis questions that we want to answer with our visual analytics system. 
These questions have led us to design the methods described in \autoref{sec:methodology}.

\begin{description}
	\item[AQ1] Do multiple models (even trained with similar ML methods) have different internal criteria for predictions and how different are they?

	An answer to this question will show the importance of interpretations and comparisons of multiple models. 
	For example, the gradient boosting method and its variations have the same theoretical background.
	We want to first know whether these methods have significant differences in their prediction criteria. 
	Afterwards, we can move on to the next level of analysis.

	\item[AQ2]
	\label{itm:AQ2}
	Which model likely has a higher consistency in its prediction criteria and should be more trusted consequently?  
    Furthermore, which range of which features are more reliable within a model's prediction criteria?

	After getting an overview of different models' consistency,	users would like to know the details of each model's consistency for each feature, 
	so that they can further select the reliable features to trust when viewing the predictions.
	Moreover, even within one feature, the consistency could vary across each range of feature values. 
	For example, a model could have high consistency in the prediction for patients who are over 60 while having low consistency for younger patients.
	Therefore, our system should also support these analyses.
\end{description}

%% file: for_arxiv_submission/4_methods.tex
\section{Methodology}
\label{sec:methodology}

We describe our dataset, prediction task, ML methods, interpretation measures, and consistency measures of interpretations.
These are used in our visual analytics system, as described in~\autoref{sec:interface}.

\subsection{Data and a Prediction Task}
\label{sec:data}

We use the MIMIC-III dataset~\cite{mimiciii}, which is a large, open-access clinical database composed of de-identified critical care unit admission records for over 40,000 patients.
The dataset includes demographics, vital sign measurements, admission information, 
test results, medications, procedures, and mortality. 

From this dataset, our prediction task is to foresee the chance of in-hospital mortality, 
given a patient's current and previous admission records. 
This large database consists of patients with more than 14,000 types of diagnoses,
and thus it is irrational to predict the status of patients with drastically different diagnoses.
To concretize our prediction task, and to make the ML models' interpretable information more reasonable,
we extract admission records that contain the same diagnosis, specifically patients diagnosed with Atrial Fibrillation (AF).
Then, we process the database into a tabular dataset, containing extracted relevant features.
As a result, we obtain 8 features from 12,886 AF patients. These features include demographic information, admission status, and information within their inpatient stay (e.g. the number of ICU stays, \emph{icustays\_num}).

Although we use a specific dataset for the development of our visualization interface, our analysis methods and visualizations are designed to be applicable to other clinical datasets.

\subsection{Machine Learning Methods}
\label{sec:ml_methods}

For our analysis, we use six different tree-based methods: decision tree (DT)~\cite{DT}, random forest (RF)~\cite{RF}, gradient boosting decision tree (GBDT)~\cite{GB}, light gradient boosting machine (LightGB)~\cite{LGB}, CatBoost~\cite{CGB}, and XGBoost~\cite{XGB}. 
These six methods are chosen for their wide usage for clinical predictions and often provide satisfying performance.

As described in~\autoref{sec:interpret_measures}, we use model-agnostic interpretation methods to understand the models' rationales. 
Therefore, although we use these six methods throughout the rest of the content of the paper, our methodology in the ensuing subsections is generic enough to apply on different ML methods, including deep learning models.

\subsection{Analysis Methods for Understanding Models' Internal Criteria for Their Predictions}
\label{sec:interpret_measures}

We measure feature contributions to compare multiple ML models' internal criteria for their predictions (\textbf{AQ1}).
Given an ML model and the classes of the prediction target, a feature contribution represents how strongly each feature affects the prediction results.
Typically, for a binary prediction task, a feature contribution can be either a positive (contribute to positive class), zero (neutral), or negative (contribute to negative class) value.
In terms of granularity, there are global and local feature contributions.
A global feature contribution represents a general effect of that feature to the overall prediction across all records whereas a local feature contribution shows an impact of each individual record of a feature to the corresponding prediction. 

To answer the first part of \textbf{AQ1}---having an overview of multiple models' inner criteria---adopting either global or local feature contributions should be sufficient enough.
However, for the second part of \textbf{AQ1} and \textbf{AQ2}, we should offer comparisons at a more detailed level.
Therefore, instead of obtaining an overview of each feature's impact on the predictions (i.e., global contributions), we have decided to measure local feature contributions. 

To obtain local feature contributions, for DT and RF, we use the method described in~\cite{Palczewska2014}, and for the other methods,
we adopt the SHAP value~\cite{SHAP}.
Between the two methods, the SHAP value is model-agnostic, and, thus, can be adapted to measure feature contributions of any other models.

Though we use tree-based ML models for this study's experiments, we still employ model-agnostic interpretation methods for two reasons. Firstly, while the theoretical backgrounds of the tree-based models are similar, each of them still employs a different technique and provides a different interpretation method. 
Using a model-agnostic method provides a fair comparison across the models. 
Second, employing such interpretation methods can help our methodology be more generalizable for other potential ML models.

Let $\bm{v}_i$ be a vector of feature values of the $i$-th data record ($i = 1, \cdots, n$). 
$\bm{v}_i$ can be represented as $\bm{v}_i = (v_{i,1}, \cdots, v_{i,j}, \cdots, v_{i,m})$ where $m$ is the number of features and $v_{i,j}$ is the $j$-th feature value of the $i$-th data record.
We obtain the local feature contributions of all features for each data record. 
Specifically, $\bm{\phi}_i = (\phi_{i,1}, \cdots, \phi_{i,j}, \cdots, \phi_{i,m})$ where $\bm{\phi}_i$ is the local feature contributions of $\bm{v}_i$ and $\phi_{i,j}$ is the $j$-th feature's contribution of the $i$-th data record. 
Because there are $k$ models, for each model, we compute a set of such local feature contributions with $m$ features for each of $n$ data records.
As a result, in total, we have $n \times k$ feature contributions vectors with length $m$. 
For example, our dataset described in \autoref{sec:data} has $n=12,886$ and $m=8$.
Also, we compare $k=6$ models described in \autoref{sec:ml_methods}.
Thus, in our case, we will get $12,886 \times 6 = 77,316$ vectors with length $8$.

However, it is difficult to review a large set of feature contributions (e.g., 77,316 vectors) one-by-one. 
Thus, to effectively obtain an answer for \textbf{AQ1}, we provide an intuitive overview of the feature contributions' similarities across the different models.
The visualized example can be found in \autoref{fig:t-SNE}.
We adopt the dimensionality reduction (DR) methods, such as t-SNE~\cite{tsne}, to project these vectors of dimension $m$ onto a 2D plot.
By using DR methods, data points with similar feature contributions will be placed close to each other.
In addition, the data points can be color-coded by their corresponding prediction models.
Therefore, by reviewing the distributions of clusters visually appeared in the DR result in conjunction with the color information of the prediction models,
the users can explore how the local feature contributions vary among multiple models. 
Refer to \autoref{sec:case_study} for the detailed analysis example.

\subsection{Consistency Measures of Models' Decision Criteria with Interpreted Information}
\label{sec:consistency_measures}
\begin{figure}[tb]
    \captionsetup{farskip=0pt}
	\centering
	\subfloat[Model A]{
     \includegraphics[width=\linewidth]{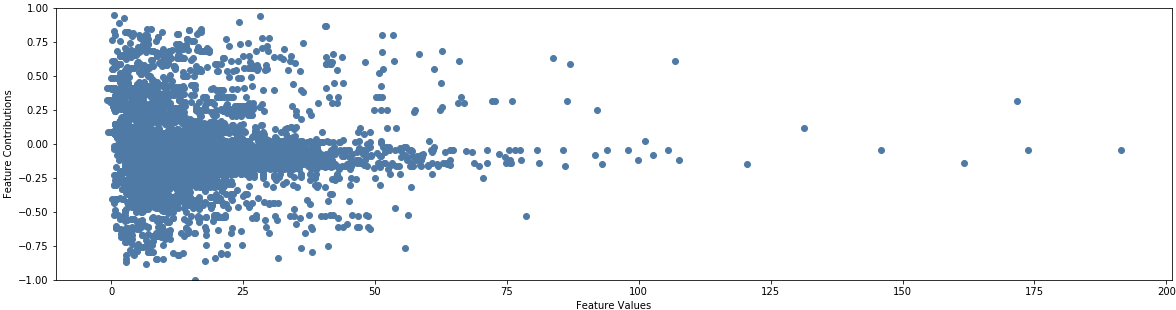}
        \label{fig:fluc_a}
    }
    \\
    \subfloat[Model B]{
     \includegraphics[width=\linewidth]{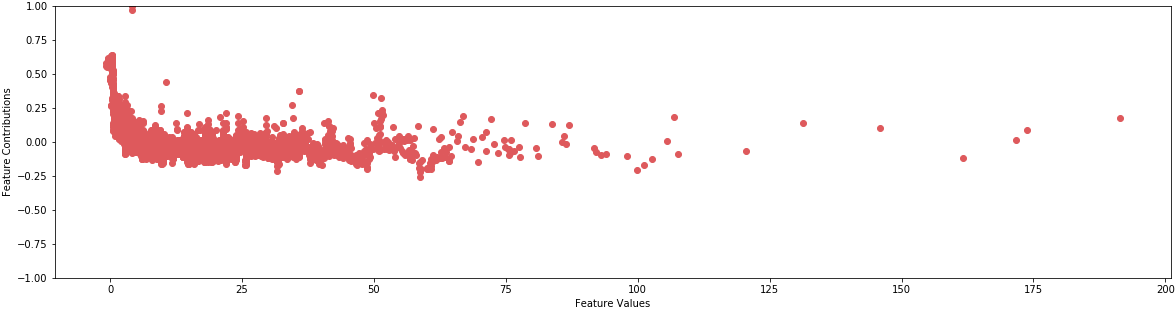}
        \label{fig:fluc_b}
    }
    \caption{The scatterplots of local feature contributions ($y$-direction) and input feature values ($x$-direction) for two ML models. Model B's prediction criteria have a higher consistency than Model A's.}
    \label{fig:fluctuation}
\end{figure}

After we compare each model's internal criteria for its predictions by using the methods described in \autoref{sec:interpret_measures}, we want to analyze the consistency of each model (\textbf{AQ2}). 
As described in \textbf{AQ2}, we consider that a model has high consistency when its criteria for the prediction is robust for a small perturbation in an input feature value. 
Since local feature contributions characterize how each feature value contributes to the prediction, we concretize the definition of consistency as follows.
\begin{description}
    \item[Consistency of the model's internal criteria:]
    The model's internal criteria have a lower consistency when local feature contributions are more independent from input feature values (the feature contributions are decided randomly regardless of feature values).
    On the other hand, the criteria have a higher consistency when local feature contributions have a higher dependency on input feature values (the feature contributions are more decisive based on feature values).
\end{description}
For example, two scatterplots in~\autoref{fig:fluctuation} visualize the local feature contributions ($y$-direction) against the feature values ($x$-direction) for two different ML models.
Here we have the same number of samples for each model (12,886 samples for each).
We consider that Model A corresponding to~\autoref{fig:fluc_a} has a lower consistency than Model B for~\autoref{fig:fluc_b}. 
For example, for the input values in a range from 0 to 25, Model A has random feature contributions. 
Thus, for such input values, Model A keeps changing how much it should rely on the corresponding feature. 
This shows Model A's low consistency in its criteria for the prediction.

With the definition above, we can obtain consistency with ``measures of dependence''~\cite{reshef2016measuring} between the input feature values and local feature contributions. 
Measures of dependence capture how strongly two variables are dependent on each other.
For example, Pearson's correlation coefficient is one of the most popular measures of dependence. 
As we can see the example in \autoref{fig:fluc_b}, feature values and local feature contributions often form non-linear dependency.
Therefore, we decide to use measures that can be used to capture both linear and non-linear dependencies. 
Moreover, it is ideal to use measures that do not have any assumption of the variables' distributions. 
The recently developed measures, such as the mutual information (MI)~\cite{Cover2006}, maximal information coefficient (MIC$_{e}$)~\cite{Reshef2011}, and total information coefficient (TIC$_{e}$)~\cite{reshef2016measuring} fulfill the above requirements.
Among these measures, TIC$_{e}$ is known for the best measure for various datasets~\cite{Reshef2018,romano2018randomized}.
We also tested these three measures on our dataset and 
TIC$_{e}$ produced the most reasonable conclusions. 
Therefore, we have decided to use TIC$_{e}$ for measuring consistency.
Comparisons of these measures are discussed in~\autoref{sec:discussion}.

%% file: for_arxiv_submission/5_interface.tex
\begin{figure*}[tb]
    \includegraphics[width=\textwidth]{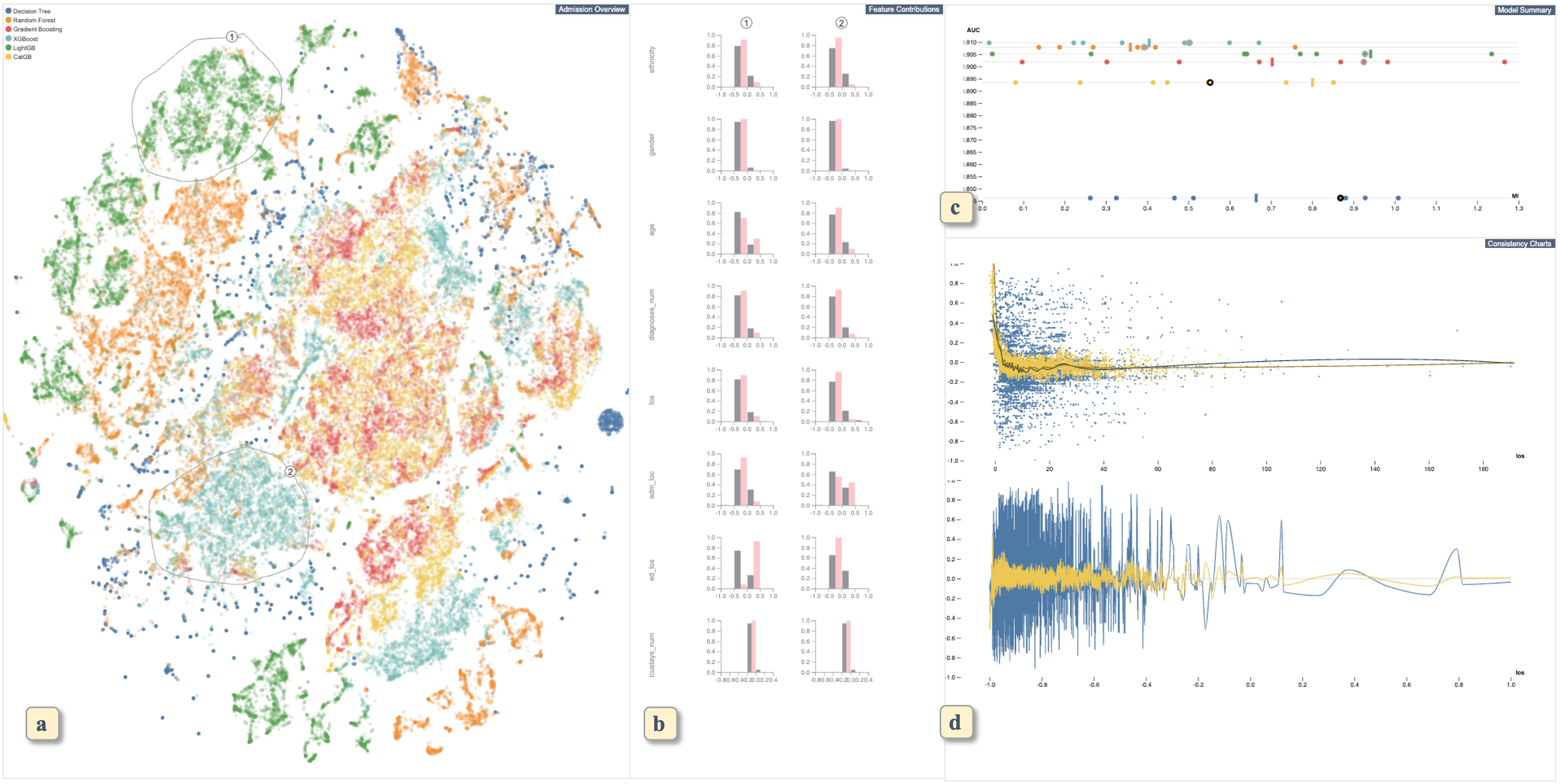}
    \captionof{figure}{A screenshot of our visual analytics system, which contains the Admission Overview (a), Feature Contributions View (b), Model Summary View (c), and Consistency Charts View (d).}
    \label{fig:teaser}
\end{figure*}

\section{Visual Analytics System}
\label{sec:interface}
We describe our visual analytics system using the methodology described in \autoref{sec:methodology}.
As shown in~\autoref{fig:teaser}, the system consists of four main views. 
The left two views, \autoref{fig:teaser}a and b, are developed for an overall comparison between different models' internal criteria of their predictions (\textbf{AQ1}); whereas the right two views, \autoref{fig:teaser}c and d, can be used for a detailed comparison of the models' consistency (\textbf{AQ2}). 
We provide a demonstration of the user interface as a supplementary video\footnote{The demonstration of our UI, \url{https://www.youtube.com/watch?v=KBZYcwEo43Q}}.

\subsection{Overall Comparison of Models' Interpretations (\autoref{fig:teaser}a and b)}

Using the method described in \autoref{sec:interpret_measures}, in \autoref{fig:teaser}a, we visualize an overview of the similarities of each model's local feature contributions, named the Admission Overview as each point represents an admission record with information of the ML model used for the prediction.
We employ t-SNE~\cite{tsne} as a dimensionality reduction (DR) method because it is suitable to find patterns (e.g., clusters) in a large dataset (77,316 data points in \autoref{fig:teaser}a). 
Specifically, we use the openTSNE~\cite{policar2019opentsne} implementation for the fast computation and precise control of t-SNE's parameters. 
We color each point based on which model it belongs to. 
We use categorical colors with enough differences in hues to distinguish from each other. 
Also, we set color transparency to be able to see overlapped points.
By viewing the points' positions for the same model and the distances among points for different models, the user can verify the diversity of different models' rationales.
For example, if two models have only a small number of overlaps, they have different prediction mechanisms (e.g., green and cyan points in \autoref{fig:teaser}a).

The Admission Overview provides a lasso selection with mouse-dragging. 
The lasso selection allows the user to select a cluster of points. 
Also, as shown in \autoref{fig:teaser}a, the user can select multiple clusters. 
The selected clusters are indicated with the drawn-lasso shapes with the identifying numbers (e.g., \textcircled{\footnotesize 1} and \textcircled{\footnotesize 2} in \autoref{fig:teaser}a).
Based on the selection, the user can review the detail differences in local feature contributions of the selected cluster(s) from the other points in the Feature Contributions view (\autoref{fig:teaser}b). 

\begin{figure}[tb]
    \centering
    \includegraphics[width=0.8\linewidth]{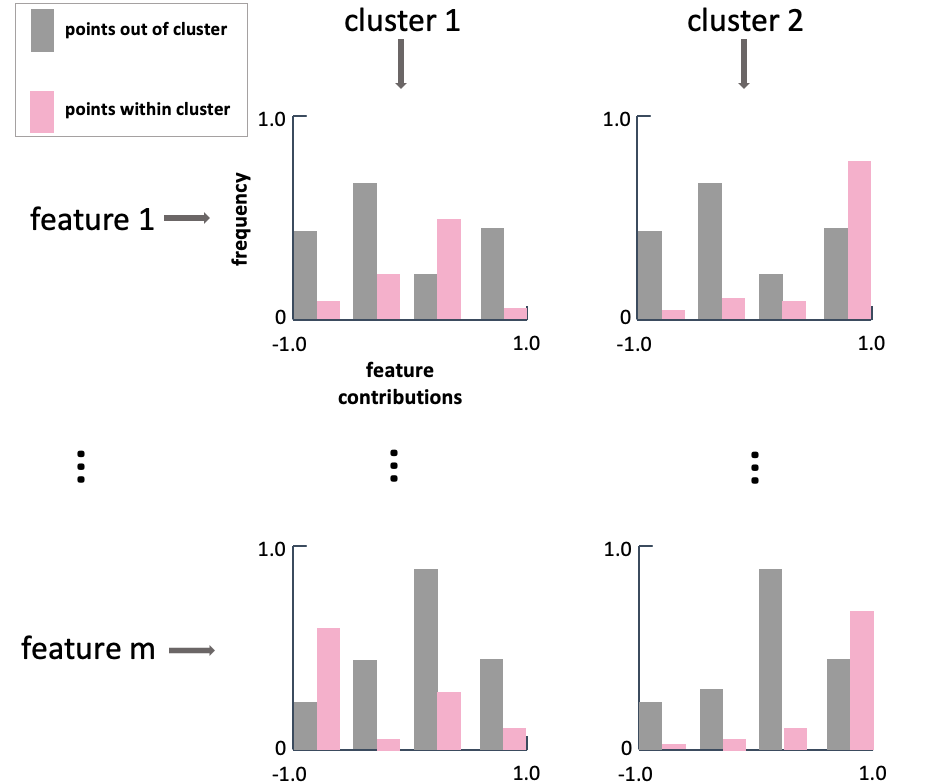}
    \caption{An illustration of Feature Contributions View's organization}
    \label{fig:fc_sketch}
\end{figure}

The Feature Contributions View in \autoref{fig:teaser}b shows a table in which cells contain histograms for the comparison of the distribution of local feature contributions.
As shown in \autoref{fig:teaser}b and \autoref{fig:fc_sketch}, each row and column correspond to a certain feature and one of selected clusters, respectively.
Then, each cell shows the distributions of the local feature contributions of the corresponding feature for the selected cluster and others.
As indicated in the legend of~\autoref{fig:fc_sketch}, the pink and gray histograms correspond to the selected cluster and others, respectively, where $y$-coordinates represent the relative frequency.
We have decided to use these two colors to differentiate the categorical colors in the Admission Overview, which are used to represent the models.
We should note that since the selected points could be members from multiple different models, we cannot assign the same color used for the model instead of pink.
By comparing the height of pink and gray bars, we can understand the differences of the selected cluster's prediction criteria from others.
For example, in \autoref{fig:fc_sketch}, cluster 2 tends to have higher feature contributions for feature 1 than compared to that of other points. 
Thus, we can say cluster 2 highly relies on feature 1 for its prediction.
Also, by comparing the histograms of each row, the user can observe which feature's distribution varies across multiple selected clusters.

Through the analysis using the Admission Overview and Feature Contribution View, users can understand which features have high contributions to the predictions of multiple selected sets of points.
Together with the information of the selected points' models, users can move on to the analysis of the consistency using \autoref{fig:teaser}c and d.

\subsection{Comparison of Models' Consistencies (\autoref{fig:teaser}c and d)}

After understanding the general differences or similarities among multiple models, we move on to the comparison of different models' consistencies in their inner rationale of decision making (\textbf{AQ2}).

We first visualize the dependencies between each feature's contributions and the values in the Model Summary View (\autoref{fig:teaser}c). 
In the scatterplots of this view, each point's $y$-coordinate represents the performance measure (accuracy rate (ACC) or area under the curve (AUC)) of the corresponding model.
The user can select one of these performance measures.
Each point's $x$-coordinate represents a value of the measure of dependency, specifically TIC$_e$ in our case. 
While ACC or AUC is the measure for each model, TIC$_e$ is the measure for each feature of each model. 
Thus, we use a horizontal line to represent each model and then, within each horizontal line, each circled dot conveys TIC$_e$ for each feature. 
Additionally, we use the rectangle shape to indicate the average TIC$_e$ of all features to show the overall consistency of the model. 
For the model information, we use the same categorical colors with the Admission Overview (\autoref{fig:teaser}a) to link the two views. 
Since the measures of prediction performance and consistency are encoded in the $x$- and $y$-coordinates respectively, it is intuitive to observe that the models whose rectangular dots are closer to the upper right corner of the plot produce more accurate, trusted results.
Across the different features for each model, the predictions that highly relies on features (circled dots) that have high TIC$_e$ can be more trusted when the corresponding model's rationales are viewed.

By hovering over each point, the user can see its detailed information (feature name, $x$, and $y$ values).
Also, all circled dots corresponding to the hovered feature will be highlighted with gray outer-rings in horizontal lines of other models for clearer comparisons.
Note that black outer-rings show the selected circled dots as explained in the description below.
The hovered example can be seen in \autoref{fig:summary}.
Through this summary view, the user can understand an overview of how different features of different models
distribute in terms of their consistencies and the prediction performance.

After obtaining the summary of the consistency of each model and each feature within a model, the Consistency Charts (\autoref{fig:teaser}d) can be used to verify the results in the Model Summary View as well as compare the consistency of different ranges of feature values.
To choose the feature and models the user wants to analyze in detail, the user can select one or multiple circled dots which correspond to a certain feature (i.e., points with the same feature name, but different colors). 

\begin{figure}[tb]
    \centering
    \includegraphics[width=\linewidth]{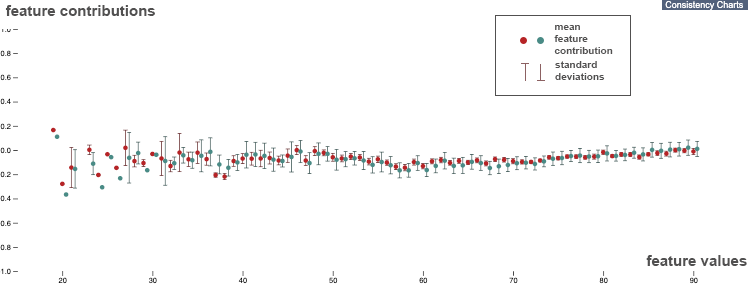}
    \caption{The Consistency Charts when visualizing categorical features}
    \label{fig:categorical}
\end{figure}

For the selected point(s), the Consistency Charts (\autoref{fig:teaser}d) provide a visual explanation of the calculated consistency in the Model Summary View as well as a more detailed inspection on the level of feature values. 
Because the features we extracted from the MIMIC-III dataset are either continuous or categorical, we provide a different visualization for each type of feature. Users can switch between different features by clicking on the points in the Model Summary View.
For continuous features, in the top view of \autoref{fig:teaser}d, we provide scatterplots of feature contributions ($y$-coordinates) against feature values ($x$-coordinates) as similar to the plots in~\autoref{fig:fluctuation}.
To be informed about the selected models, we use the same categorical color with the Admission Overview and the Model Summary View.
For each model's scatterplots, we also provide a regression line that shows how the feature contributions change in relation to the values. 
Since the relationship between feature contributions and feature values is often non-linear, we adopt LOESS regression~\cite{loess}, which is a widely used non-linear regression method.
Then, in the bottom view of \autoref{fig:teaser}d, we also plot the residuals of the regression using $y$-coordinates. 
$x$-coordinates represent feature values as similar to the top view.
This plot can assist the user to quantitatively measure how well the feature contributions behave with the distribution of feature values.
Comparisons between models can be done by overlapping the scatter plot and the residual plot of each model.
For categorical features, we provide a plot of points representing mean values and error bars of the local feature contributions for each feature value, as illustrated in~\autoref{fig:categorical}.
In order to make comparisons easier, we plot different models' points and error bars with a small gap between each of $x$-coordinates instead of simply overlaying them using the same $x$-coordinates.

By reviewing how widely points are distributed along the $y$-direction for each $x$-coordinate, the user can understand which model and/or which range of feature values has higher consistency. 
For example, in \autoref{fig:teaser}d, we can see that the yellow model generally has higher consistency than the blue model. 
Furthermore, within the yellow model, when the feature values are smaller, the residuals tend to have higher absolute values. 
Thus, when the yellow model predicts the result for the patients who have low values for this feature, the model's prediction criteria have low consistency. 

%% file: for_arxiv_submission/6_case_studies.tex
\section{Case Study}
\label{sec:case_study}

Using the preprocessed MIMIC-III dataset (refer to \autoref{sec:data}), we compare the six ML models described in \autoref{sec:ml_methods} in terms of their prediction performances, internal rationales for predictions (\textbf{AQ1}), and their consistencies (\textbf{AQ2}).

\subsection{Models' Prediction Performances}
As described in \autoref{sec:ml_methods}, we trained six models with DT, RF, GBDT, LightGBM, CatBoost, and XGBoost. 
We then obtained area under the curve (AUC) and accuracy rate (ACC) for each model, as shown in \autoref{tab:performances}.
From \autoref{tab:performances}, in terms of prediction performance, we can say that XGBoost has the best performance while other methods excluding DT have similar performance with XGBoost.

\begin{table}[h]
    \caption{Performance of different tree-based methods for prediction label ``in-hospital survival'' or ``in-hospital death''}
    \label{tab:performances}
    \scriptsize
	\centering
    \begin{tabu}{
	r
	*{7}{c}
	*{2}{r}
	}
	\toprule
	\textbf{Method} & \textbf{AUC} & \textbf{ACC} \\
	\midrule
	DT & 0.845935282 & 0.846111720 \\
	RF & 0.908171950 & 0.907813070 \\
	GBDT & 0.902461079 & 0.901788974 \\
	LightGBM & 0.894114838 & 0.893391749 \\
	CatBoost & 0.905595369 & 0.905074845 \\
	XGBoost & 0.910045645 & 0.909638554 \\
	\bottomrule
	\end{tabu}
\end{table}

\begin{figure}[tb]
    \includegraphics[width=1.0\linewidth]{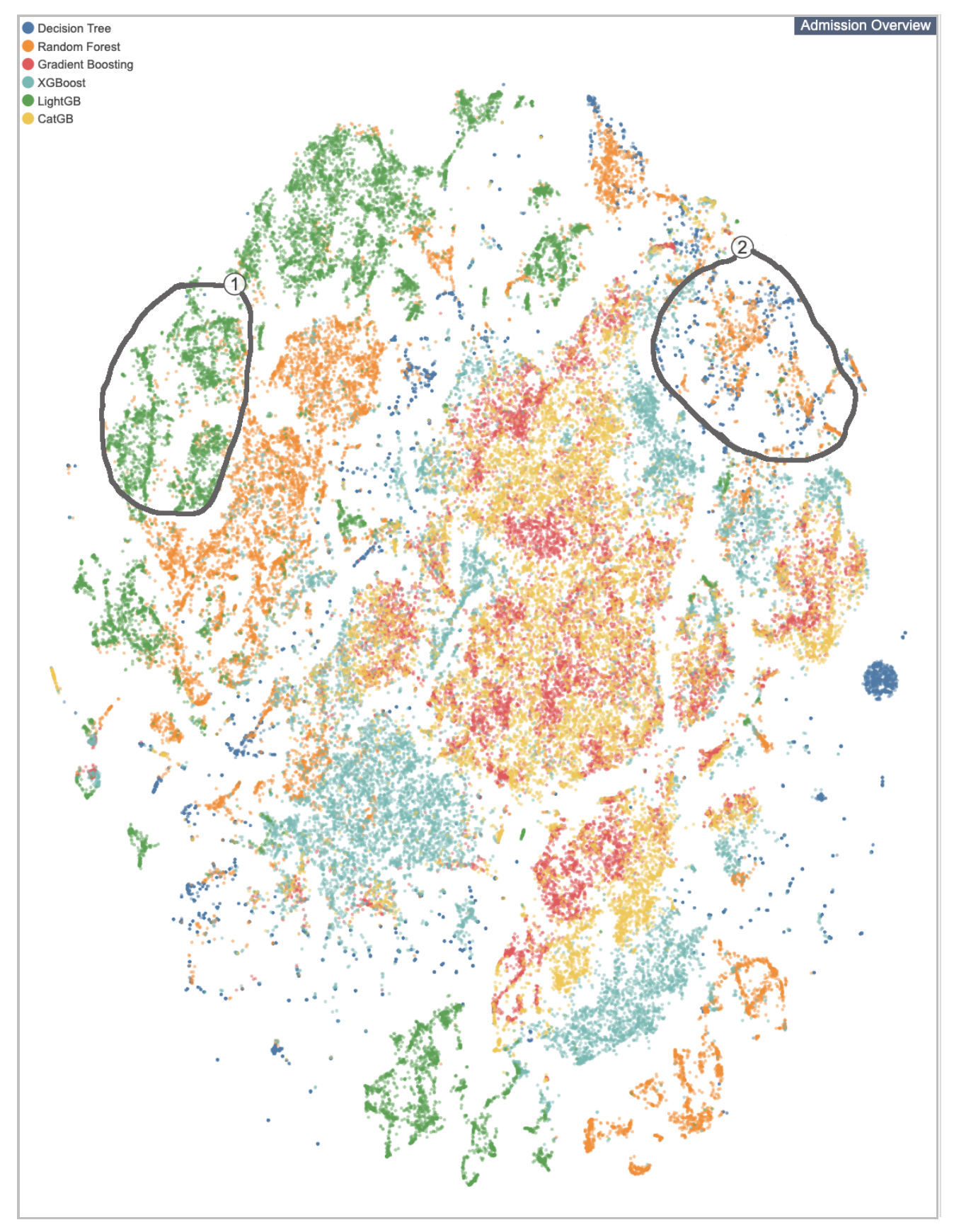}
    \caption{Feature contributions' clusters for six tree-based models}
    \label{fig:t-SNE}
\end{figure}

\subsection{Overall Comparison of Models' Internal Prediction Criteria (\textbf{AQ1})}
\label{sec:overall_comparison}

After standardizing different models' local feature contributions for each patient, we performed t-SNE on these feature contributions.
The t-SNE plot was then color-encoded and shown in the Admission Overview, as demonstrated in~\autoref{fig:t-SNE}.

From the overview of similarities of local feature contributions, we can observe a divergence in the positions of the points representing the feature contributions, 
whereas points that belong to the same models tend to be more clustered together.
However, there are exceptions for such cases. 
For example, the points of models CatGB (yellow) and Gradient Boosting (red) have many overlaps and no distinguishable boundaries,
although some areas are denser with points of one model than another.
This implies that these two models seem to share more similar prediction criteria compared to others. 
However, we can find two general tendencies from \autoref{fig:t-SNE}.
First, most of these six models' inner criteria tend to have differences from each other as we can see distinct clusters for each model. 
Second, even within the same model, each model tends to have different prediction criteria based on patients' features.
For example, for LightGB (green), while there are several clusters around the top left of \autoref{fig:t-SNE}, we can see a distinct cluster around the bottom center.

\begin{figure}[tb]
	\centering
    \includegraphics[width=0.5\linewidth]{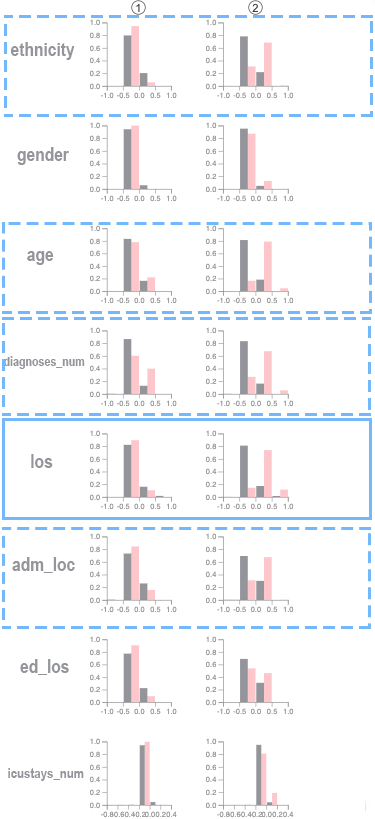}
    \caption{Feature contributions' distributions of selected clusters}
    \label{fig:fc_view}
\end{figure}

Then, we select two clusters of points within the overview to review the differences of their local feature contributions in detail.
As seen in \autoref{fig:t-SNE}, Cluster 1 contains mostly LightGB points, while Cluster 2 contains mostly DT and RF points.
The Feature Contributions View visualizes the local feature contributions' distribution of these two sets of points, as shown in~\autoref{fig:fc_view}.
By comparing each feature within the same row, we discover that there are significant differences for features \emph{ethnicity}, \emph{age}, \emph{diagnoses\_num} (number of diagnoses), \emph{los} (length of stay), and \emph{adm\_loc} (admission location), as highlighted in~\autoref{fig:fc_view}.
In the following subsection, we use feature \emph{los} as an example to demonstrate the use cases of the remaining views.

\begin{figure}[tb]
	\centering
    \includegraphics[width=1.0\linewidth]{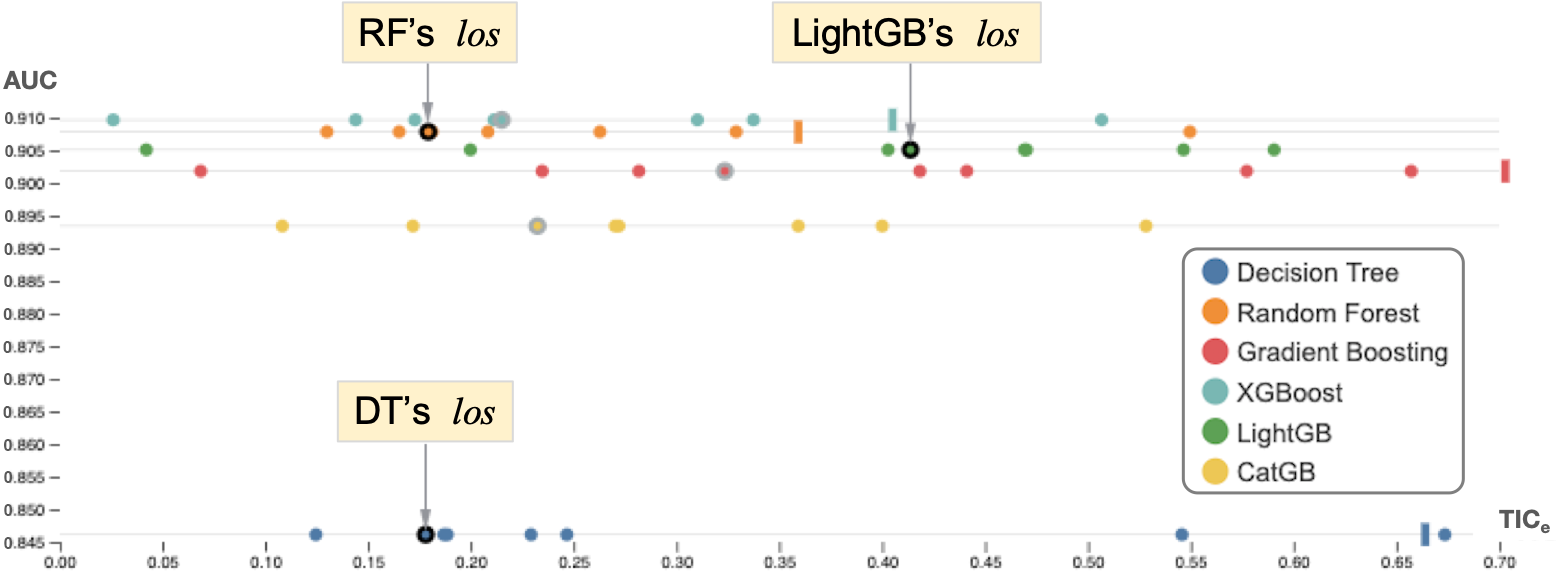}
    \caption{Overview of each feature's consistency values for each model}
    \label{fig:summary}
\end{figure}

\subsection{Comparison of Models' Consistencies (\textbf{AQ2})}

We then move on to the Model Summary view to analyze the differences in consistencies of the feature \emph{los} across models.
Here, we focus on the three models of which local feature contributions are included in the selected two clusters (i.e., LightGB in Cluster 1; DT and RF in Cluster 2).
As shown in~\autoref{fig:summary}, the highlighted points correspond to feature \emph{los} of the three models.
As observed in~\autoref{fig:summary}, LightGB's \emph{los} feature has a much higher TIC$_e$ than DT's and RF's. 
Therefore, it can be inferred that clinicians can rely more on \emph{los} when they adopt LightGB model while they cannot rely as much on \emph{los} when using DT or RF.

\begin{figure}[tb]
	\centering
    \includegraphics[width=\linewidth]{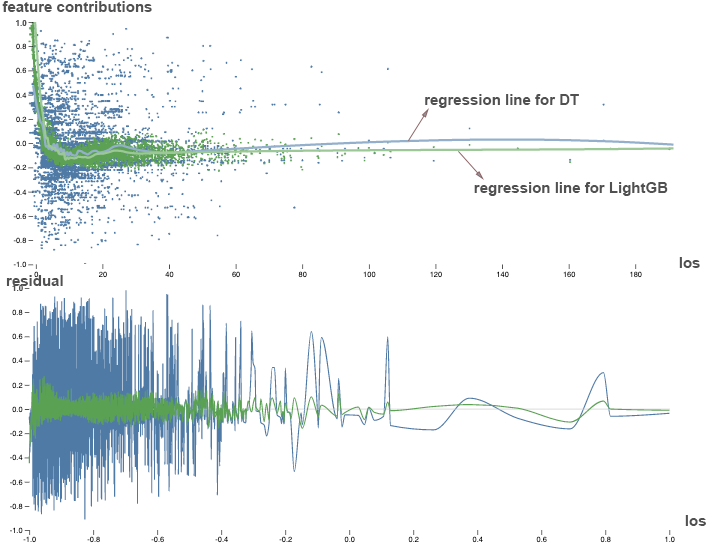}
    \caption{Detailed consistency comparison for feature \emph{los} of models DT (blue) and LightGB (green)}
    \label{fig:cc_view}
\end{figure}

By clicking the corresponding points for feature \emph{los} of LightGB and DT as an example, we analyze the detailed differences of their consistencies with the Consistency Charts, as shown in~\autoref{fig:cc_view}.
Through the comparisons between the overlaying scatterplots, we can see that it is intuitive and convincing that LightGB has more consistency than DT because the points tend to be more closely distributed around the regression line.
Furthermore, when looking at the residual plots, as shown in \autoref{fig:cc_view}, we observe that DT's residuals tend to have a wider range for small values of \emph{los}, whereas LightGB's tend to have small residuals for any values of \emph{los}.
Through this view, we can say that LightGB's trends of \emph{los}'s feature contributions can be more trusted.

Through this case study, we demonstrate how effectively we can answer \textbf{AQ1} and \textbf{AQ2} with our visual analytics system. 
Here we have only shown a certain flow of analysis. 
However, we can try various different sets of selections and gain more comprehensive insights.
For example, the user would also want to select and analyze different points in \autoref{fig:t-SNE}.  
Such analysis and exploration can be performed with the flexible interactions supported in our system.  

%% file: for_arxiv_submission/7_discussion.tex
\section{Discussion}
\label{sec:discussion}

We discuss our algorithm choice and visualization design; then the limitations of our methods and future work.

\subsection{Dependency Measures}
During the selection process of algorithms, we tried several measures of dependency between two random variables to characterize consistency.
Since we use the Consistency Charts to verify the calculated dependency values, we also performed comparisons of the dependency measures with this view.
As stated in \autoref{sec:consistency_measures}, MI, MIC$_e$, and TIC$_e$ could be used to measure the dependency between the feature contributions and feature values.
Thus, we tried all these measures and viewed the relationships between the computed dependencies and visual results in the Consistency Charts.

\begin{figure}[tb]
    \captionsetup{farskip=0pt}
    \centering
	\subfloat[MI]{
     \includegraphics[width=\linewidth]{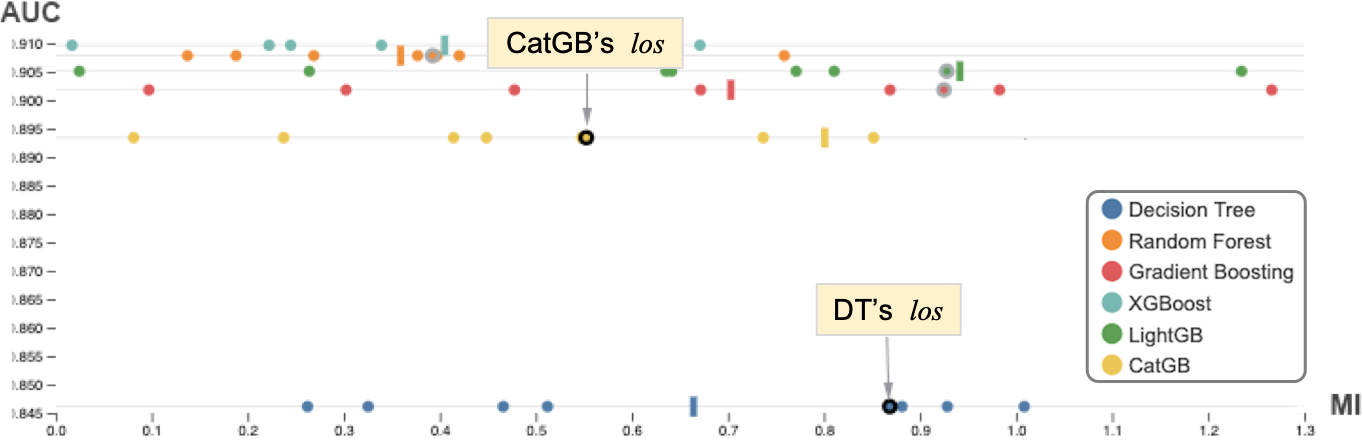}
        \label{fig:dep_sub1}
    }
    \\
    \subfloat[MIC$_e$]{
     \includegraphics[width=\linewidth]{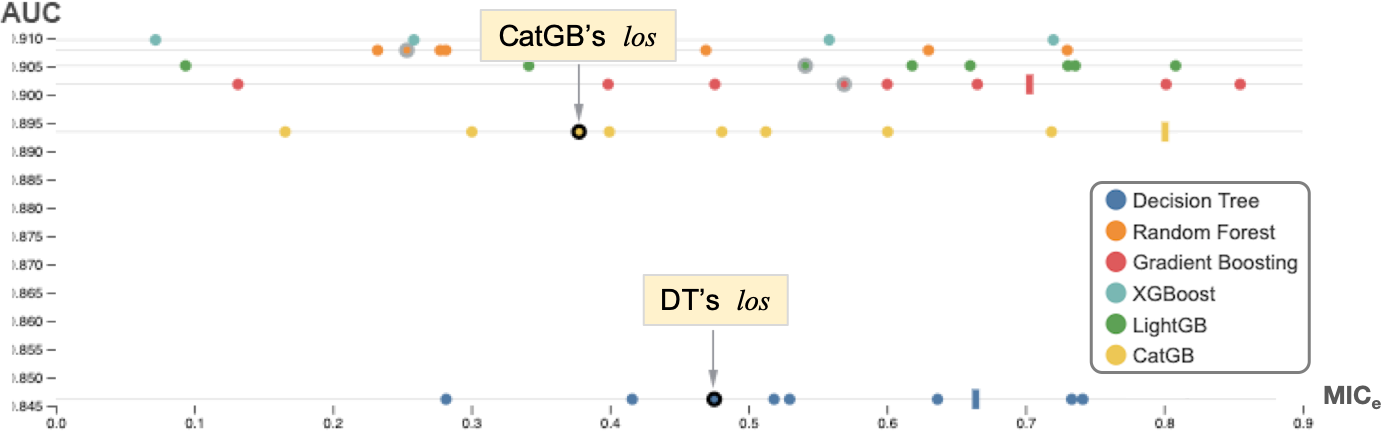}
        \label{fig:dep_sub2}
    }
    \vspace{-7pt}
    \caption{The screenshots of the Model Summary View when using (a) MI and (b) MIC$_e$ as dependency measures. In both plots, feature \emph{los} of DT has a higher dependency value than the same feature of CatGB.}
    \label{fig:dep_compare}
\end{figure}

\begin{figure}[tb]
    \centering
    \includegraphics[width=0.98\linewidth]{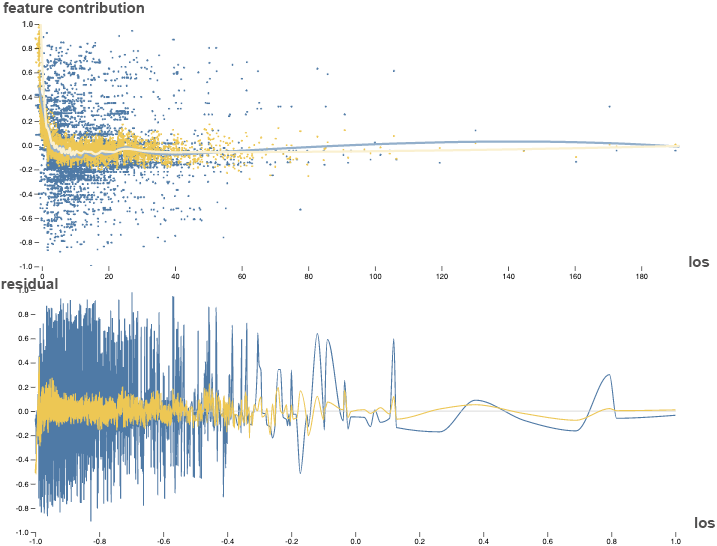}
    \caption{The Consistency Charts of LightGB (yellow) and DT (blue).}
    \label{fig:consistency}
\end{figure}

Through this comparison, we observed that although the three measures provided reasonable results for most of the features,
MI and MIC$_e$ have capricious behaviors when evaluating
dependencies, especially when there are large ranges of feature contributions for the same feature values (i.e., widely different $y$-coordinates for the same $x$-coordinates in the Consistency Charts).
On the other hand, TIC$_e$ is more stable for any types of features
and produces more rational results.
For example, as shown in~\autoref{fig:dep_compare}, 
both MI (\autoref{fig:dep_sub1}) and MIC$_e$ (\autoref{fig:dep_sub2}) indicate that feature \emph{los} of model DT has a high dependency value, even higher than CatGB. 
However, as shown in \autoref{fig:consistency}, by reviewing the detailed consistency information, we noticed that CatGB (yellow) should have a higher dependency value than DT (blue). This is because their range of feature contributions over similar feature values are relatively smaller than DT, and their residuals are closer to 0.
By adopting TIC$_e$ as the dependency measure, the unexpected behavior of the previous two measures has been solved, as we have already shown in \autoref{fig:summary}. 
Thus, we have chosen TIC$_e$ for our consistency measure.
This example from our experiment shows that both the analysis algorithm and related visualization can be evaluated by coupling with each other and comparing them; as a result, we can select a better algorithm and/or visualization.

\subsection{Visualizations for Categorical Features}
Similar to continuous features, for categorical features, we first designed the Consistency Charts using both scatterplots of distributions (e.g., the top view of \autoref{fig:consistency}) and an assisting chart that shows how much local feature contributions differ at the same feature value (e.g., the bottom view of \autoref{fig:consistency}).
For categorical features, the counterpart of the regression line of continuous features is the mean or median value for each category; the equivalence of residuals is the error bars calculated around the mean or median (e.g., bars showing standard deviations).
However, the mean or median and error bars are usually plotted together.
Therefore, instead of showing them in two different views, we first decided to follow this common way (i.e., showing both in one view).

\begin{figure}
    \captionsetup{farskip=0pt}
	\centering
    \includegraphics[width=\linewidth]{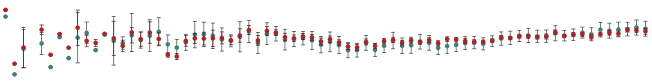}
    \caption{Another option we tried for the visualization of categorical features in the Consistency Charts.}
    \label{fig:choice_2}
\end{figure}

However, as shown in \autoref{fig:choice_2}, by following this format, the visualized results suffered from occlusion and cluttering.
This happened because the points (red and teal colored dots) and error bars shared the same $x$-coordinate.
Because the categorical values usually would not take many different values (e.g., about 70 in \autoref{fig:choice_2}), we have enough space to use slightly different $x$-coordinates around each corresponding categorical value.
Thus, as shown in \autoref{fig:categorical}, we tried a plot placing different models' points and error bars with a small gap in $x$-coordinates.
In this way, we were able to view and compare the means and error bars for different models more clearly.
Therefore, we decided to use this design for our visualization.

\subsection{Limitations}
\noindent\textbf{Scalability for the number of features and ML models.}
Our visualizations provide enough scalability for the number of data records (e.g., patients). 
For example, the Admission Overview employs t-SNE~\cite{tsne} for dimensionality reduction and can visualize the overview of similarity of local feature contributions even for tens of thousands of points. 
However, for the number of features and the number of ML models, our visualizations have limited scalability.
The number of ML models we can support is limited because we use colors to indicate the corresponding ML model. 
Thus, our visualizations can deal with less than about 10 models. 
We can address this problem by aggregating multiple models based on their similarities in a certain aspect. 
For example, as shown in \autoref{sec:overall_comparison}, CatGB and Gradient Boosting have similar distributions of local feature contributions and, thus, the user might want to analyze them as one aggregated model.

For the number of features, we would need to improve several designs if the data were to contain many features. 
For instance, the Feature Contributions View in \autoref{fig:teaser}b shows all features' information by aligning them for each row.
This way is reasonable for our dataset consisting of 8 features. 
However, when there are more than 10 features, showing and analyzing all features' information is not realistic.
In this case, the system should automatically suggest which features the user should review to understand the differences between the selected points from the Admission Overview (\autoref{fig:teaser}a) and others. 
For example, we can support such functionality by using the method introduced in \cite{fujiwara2020ccpca}.

\vspace{2pt}
\noindent\textbf{Variety of analyzed ML methods.} 
In the case study of our clinical data predictions, we adopted the tree-based ML methods. 
In terms of generalization, one future direction is to develop heuristic comparison methods for any kind of ML methods, including deep learning methods. 
For example, to extend our analysis on time series prediction methods,
besides the reliability of each feature, we would also like to compare consistencies of different time steps.
For deep learning methods, the reliability of the last few layers' outputs, for instance, should also be our focus of comparisons.

\vspace{2pt}
\noindent\textbf{Supported Analyses.} 
Our methods and visualizations can help understand many aspects of the ML models' prediction rationales.
However, there are still some points that cannot be explained by the current system.
For example, while we can analyze how each model relies on some features more than the other features as their learned results, we cannot know how each model obtained such criteria.
More specifically, as stated in \autoref{sec:overall_comparison}, through the analysis, we found that CatGB and Gradient Boosting tend to have similar local feature contributions.
However, we cannot further analyze why these two methods reached such results. 
Therefore, our work needs to be extended to cover such exceptions.
For such analyses, we can incorporate the methods developed for understanding the learning process of ML models. 
For example, for the tree-based ML methods, we can use the method by Liu et al.~\cite{liu2017visual}.

%% file: for_arxiv_submission/8_conclusions.tex
\section{Conclusions}
We have developed a visual analytics system that utilizes quantitative methods to observe and compare multiple models' reliability through their interpretable information.
Using our system, insights of multiple models' internal criteria can be obtained and their reliability can be further evaluated on both overview and individual feature levels.
Through the case study, we have demonstrated the usefulness of this visual analytics to aid clinical researchers' model selections.
Our visual analytics system can be extended to have additional support for various machine learning methods and a more scalable interface that provides functionalities for a more comprehensive analysis.

%% file: 00_main_for_arxiv_submission.bbl
\begin{thebibliography}{10}

\bibitem{adadi2018peeking}
A.~Adadi and M.~Berrada.
\newblock Peeking inside the black-box: A survey on explainable artificial
  intelligence ({XAI}).
\newblock {\em IEEE Access}, 6:52138--52160, 2018.

\bibitem{Bousquet2002}
O.~Bousquet and A.~Elisseeff.
\newblock Stability and generalization.
\newblock {\em J. Machine Learning Research}, 2(Mar):499--526, 2002.

\bibitem{caruana2015intelligible}
R.~Caruana, Y.~Lou, J.~Gehrke, P.~Koch, M.~Sturm, and N.~Elhadad.
\newblock Intelligible models for healthcare: Predicting pneumonia risk and
  hospital 30-day readmission.
\newblock In {\em Proc. SIGKDD}, pp. 1721--1730, 2015.

\bibitem{missing}
Z.~Che, S.~Purushotham, K.~Cho, D.~Sontag, and Y.~Liu.
\newblock Recurrent neural networks for multivariate time series with missing
  values.
\newblock {\em Scientific Reports (Nature Publisher Group)}, 8:1--12, 04 2018.

\bibitem{XGB}
T.~Chen and C.~Guestrin.
\newblock {XGBoost}: A scalable tree boosting system.
\newblock In {\em Proc. SIGKDD}, pp. 785--794, 2016.

\bibitem{Retain}
E.~Choi, M.~T. Bahadori, J.~Sun, J.~Kulas, A.~Schuetz, and W.~Stewart.
\newblock {RETAIN}: An interpretable predictive model for healthcare using
  reverse time attention mechanism.
\newblock In {\em Proc. NIPS}, pp. 3504--3512, 2016.

\bibitem{loess}
W.~S. Cleveland.
\newblock Robust locally weighted regression and smoothing scatterplots.
\newblock {\em J. American Statistical Association}, 74(368):829--836, 1979.

\bibitem{instance}
D.~Collaris, L.~M. Vink, and J.~J. van Wijk.
\newblock Instance-level explanations for fraud detection: {A} case study.
\newblock {\em CoRR}, abs/1806.07129, 2018.

\bibitem{Cover2006}
T.~M. Cover and J.~A. Thomas.
\newblock {\em Elements of Information Theory (Wiley Series in
  Telecommunications and Signal Processing)}.
\newblock Wiley-Interscience, New York, NY, USA, 2006.

\bibitem{Devroye1991}
L.~Devroye.
\newblock Exponential inequalities in nonparametric estimation.
\newblock In {\em Nonparametric functional estimation and related topics}, pp.
  31--44. Springer, 1991.

\bibitem{Devroye1979}
L.~{Devroye} and T.~{Wagner}.
\newblock Distribution-free performance bounds for potential function rules.
\newblock {\em IEEE Trans. on Information Theory}, 25(5):601--604, Sep. 1979.

\bibitem{CGB}
A.~V. Dorogush, V.~Ershov, and A.~Gulin.
\newblock {CatBoost}: gradient boosting with categorical features support.
\newblock {\em CoRR}, abs/1810.11363, 2018.

\bibitem{tree-based3}
M.~Fouad, M.~Mohamoud, M.~Hagag, and A.~Akl.
\newblock Prediction of long term living donor kidney graft outcome: Comparison
  between rule based, decision tree and linear regression.
\newblock {\em Int. J. Advanced Research in Computer Science}, 3:185, 04 2015.

\bibitem{GB}
J.~H. Friedman.
\newblock Greedy function approximation: A gradient boosting machine.
\newblock {\em The Annals of Statistics}, 29(5):1189--1232, 2001.

\bibitem{fujiwara2020ccpca}
T.~Fujiwara, O.-H. Kwon, and K.-L. Ma.
\newblock Supporting analysis of dimensionality reduction results with
  contrastive learning.
\newblock {\em IEEE Trans. on Visualization and Computer Graphics},
  26(1):45--55, 2020.

\bibitem{tree-based2}
A.~S. Goldfarb-Rumyantzev, J.~D. Scandling, L.~Pappas, R.~J. Smout, and
  S.~Horn.
\newblock Prediction of 3-yr cadaveric graft survival based on pre-transplant
  variables in a large national dataset.
\newblock {\em Clinical Transplantation}, 17(6):485--497, 2003.

\bibitem{VisProgression}
S.~{Guo}, Z.~{Jin}, D.~{Gotz}, F.~{Du}, H.~{Zha}, and N.~{Cao}.
\newblock Visual progression analysis of event sequence data.
\newblock {\em IEEE Trans. on Visualization and Computer Graphics},
  25(1):417--426, Jan 2019.

\bibitem{Haussler1995}
D.~Haussler.
\newblock Part 1: Overview of the probably approximately correct ({PAC})
  learning framework.
\newblock \url{http://web.cs.iastate.edu/~honavar/pac.pdf}, 1995.
\newblock Accessed: 2019-12-19.

\bibitem{CarePre}
Z.~Jin, S.~Cui, S.~Guo, D.~Gotz, J.~Sun, and N.~Cao.
\newblock {CarePre}: An intelligent clinical decision assistance system.
\newblock {\em ACM Trans. on Computing for Healthcare}, 2019.

\bibitem{mimiciii}
A.~E. Johnson, T.~J. Pollard, L.~Shen, H.~L. Li-wei, M.~Feng, M.~Ghassemi,
  B.~Moody, P.~Szolovits, L.~A. Celi, and R.~G. Mark.
\newblock {MIMIC-III}, a freely accessible critical care database.
\newblock {\em Scientific Data}, 3:160035, 2016.

\bibitem{kawaler2012}
E.~Kawaler, A.~Cobian, P.~Peissig, D.~Cross, S.~Yale, and M.~Craven.
\newblock Learning to predict post-hospitalization {VTE} risk from {EHR} data.
\newblock In {\em AMIA Annual Symp. Proc.}, vol. 2012, p. 436. American Medical
  Informatics Association.

\bibitem{LGB}
G.~Ke, Q.~Meng, T.~Finley, T.~Wang, W.~Chen, W.~Ma, Q.~Ye, and T.-Y. Liu.
\newblock {LightGBM}: A highly efficient gradient boosting decision tree.
\newblock In {\em Proc. NIPS}, 2017.

\bibitem{Kearns1999}
M.~{Kearns} and D.~{Ron}.
\newblock Algorithmic stability and sanity-check bounds for leave-one-out
  cross-validation.
\newblock {\em Neural Computation}, 11(6):1427--1453, Aug 1999.

\bibitem{Kearns1994}
M.~J. Kearns and U.~V. Vazirani.
\newblock {\em An Introduction to Computational Learning Theory}.
\newblock MIT Press, 1994.

\bibitem{mortalitypred}
J.~L. Koyner, K.~A. Carey, D.~P. Edelson, and M.~M. Churpek.
\newblock The development of a machine learning inpatient acute kidney injury
  prediction model.
\newblock {\em Critical Care Medicine}, 46(7):1070—1077, July 2018.

\bibitem{RetainVis}
B.~C. Kwon, M.-J. Choi, J.~T. Kim, E.~Choi, Y.~B. Kim, S.~Kwon, J.~Sun, and
  J.~Choo.
\newblock {RetainVis}: Visual analytics with interpretable and interactive
  recurrent neural networks on electronic medical records.
\newblock {\em IEEE Trans. on Visualization and Computer Graphics},
  25(1):299--309, 2018.

\bibitem{LSTM}
Z.~C. Lipton, D.~C. Kale, C.~Elkan, and R.~C. Wetzel.
\newblock Learning to diagnose with {LSTM} recurrent neural networks.
\newblock In {\em Proc. Int. Conf. on Learning Representations}, 2016.

\bibitem{LiuTowards2017}
S.~Liu, X.~Wang, M.~Liu, and J.~Zhu.
\newblock Towards better analysis of machine learning models: A visual
  analytics perspective.
\newblock {\em Visual Informatics}, 1(1):48 -- 56, 2017.

\bibitem{liu2017visual}
S.~Liu, J.~Xiao, J.~Liu, X.~Wang, J.~Wu, and J.~Zhu.
\newblock Visual diagnosis of tree boosting methods.
\newblock {\em IEEE Trans. on Visualization and Computer Graphics},
  24(1):163--173, 2017.

\bibitem{Lugosi1994}
G.~{Lugosi} and M.~{Pawlak}.
\newblock On the posterior-probability estimate of the error rate of
  nonparametric classification rules.
\newblock {\em IEEE Trans. on Information Theory}, 40(2):475--481, March 1994.

\bibitem{SHAP}
S.~M. Lundberg and S.-I. Lee.
\newblock A unified approach to interpreting model predictions.
\newblock In {\em Proc. NIPS}, pp. 4765--4774, 2017.

\bibitem{MingRuleMatrix}
Y.~{Ming}, H.~{Qu}, and E.~{Bertini}.
\newblock {RuleMatrix}: Visualizing and understanding classifiers with rules.
\newblock {\em IEEE Trans. on Visualization and Computer Graphics},
  25(1):342--352, Jan 2019.

\bibitem{TreePOD}
T.~{Mühlbacher}, L.~{Linhardt}, T.~{Möller}, and H.~{Piringer}.
\newblock {TreePOD}: Sensitivity-aware selection of pareto-optimal decision
  trees.
\newblock {\em IEEE Trans. on Visualization and Computer Graphics},
  24(1):174--183, Jan 2018.

\bibitem{Palczewska2014}
A.~Palczewska, J.~Palczewski, R.~Marchese~Robinson, and D.~Neagu.
\newblock {\em Interpreting random forest classification models using a feature
  contribution method}, pp. 193--218.
\newblock Springer, Cham, 2014.

\bibitem{Pearson1895}
K.~Pearson and F.~Galton.
\newblock Vii. {Note} on regression and inheritance in the case of two parents.
\newblock {\em Proc. Royal Society of London}, 58(347-352):240--242, 1895.

\bibitem{policar2019opentsne}
P.~G. Policar, M.~Strazar, and B.~Zupan.
\newblock {openTSNE}: a modular python library for t-{SNE} dimensionality
  reduction and embedding.
\newblock {\em BioRxiv}, p. 731877, 2019.

\bibitem{DT}
J.~R. Quinlan.
\newblock Learning efficient classification procedures and their application to
  chess end games.
\newblock In {\em Machine Learning: An Artificial Intelligence Approach}, pp.
  463--482. Springer, 1983.

\bibitem{Reshef2011}
D.~N. Reshef, Y.~A. Reshef, H.~K. Finucane, S.~R. Grossman, G.~McVean, P.~J.
  Turnbaugh, E.~S. Lander, M.~Mitzenmacher, and P.~C. Sabeti.
\newblock Detecting novel associations in large data sets.
\newblock {\em Science}, 334(6062):1518--1524, 2011.

\bibitem{Reshef2018}
D.~N. Reshef, Y.~A. Reshef, P.~C. Sabeti, and M.~Mitzenmacher.
\newblock An empirical study of the maximal and total information coefficients
  and leading measures of dependence.
\newblock {\em Ann. Appl. Stat.}, 12(1):123--155, 03 2018.

\bibitem{reshef2016measuring}
Y.~A. Reshef, D.~N. Reshef, H.~K. Finucane, P.~C. Sabeti, and M.~Mitzenmacher.
\newblock Measuring dependence powerfully and equitably.
\newblock {\em J. Machine Learning Research}, 17(1):7406--7468, 2016.

\bibitem{LIME}
M.~T. Ribeiro, S.~Singh, and C.~Guestrin.
\newblock "{Why} should i trust you?": Explaining the predictions of any
  classifier.
\newblock In {\em Proc. SIGKDD}, pp. 1135--1144, 2016.

\bibitem{romano2018randomized}
S.~Romano, N.~X. Vinh, K.~Verspoor, and J.~Bailey.
\newblock The randomized information coefficient: Assessing dependencies in
  noisy data.
\newblock {\em Machine Learning}, 107(3):509--549, 2018.

\bibitem{treeinterpreter}
A.~Saabas.
\newblock {TreeInterpreter}.
\newblock \url{https://github.com/andosa/treeinterpreter}.
\newblock Accessed: 2019-12-19.

\bibitem{tree-based1}
T.~Shaikhina, D.~Lowe, S.~Daga, D.~Briggs, R.~Higgins, and N.~Khovanova.
\newblock Decision tree and random forest models for outcome prediction in
  antibody incompatible kidney transplantation.
\newblock {\em Biomedical Signal Processing and Control}, 2017.

\bibitem{CDSS}
E.~H. Shortliffe and M.~J. Sepúlveda.
\newblock Clinical decision support in the era of artificial intelligence
  clinical decision support in the era of artificial intelligence clinical
  decision support in the era of artificial intelligence.
\newblock {\em JAMA}, 320(21):2199--2200, 12 2018.

\bibitem{Spearman1987}
C.~Spearman.
\newblock The proof and measurement of association between two things.
\newblock {\em The American J. Psychology}, 100(3/4):441--471, 1987.

\bibitem{RF}
{Tin Kam Ho}.
\newblock Random decision forests.
\newblock In {\em Proc. Int. Conf. on Document Analysis and Recognition},
  vol.~1, pp. 278--282 vol.1, Aug 1995.

\bibitem{tsne}
L.~van~der Maaten and G.~Hinton.
\newblock Viualizing data using t-{SNE}.
\newblock {\em J. Machine Learning Research}, 9:2579--2605, 11 2008.

\bibitem{FeatureAttr}
C.~Wang, T.~Onishi, K.~Nemoto, and K.-L. Ma.
\newblock Visual reasoning of feature attribution with deep recurrent neural
  networks.
\newblock In {\em Proc. IEEE Int. Conf. on Big Data}, pp. 1661--1668, 2018.

\bibitem{WangDeepVID}
J.~{Wang}, L.~{Gou}, W.~{Zhang}, H.~{Yang}, and H.~{Shen}.
\newblock {DeepVID}: Deep visual interpretation and diagnosis for image
  classifiers via knowledge distillation.
\newblock {\em IEEE Trans. on Visualization and Computer Graphics},
  25(6):2168--2180, June 2019.

\bibitem{zhang2018manifold}
J.~Zhang, Y.~Wang, P.~Molino, L.~Li, and D.~S. Ebert.
\newblock Manifold: A model-agnostic framework for interpretation and diagnosis
  of machine learning models.
\newblock {\em IEEE Trans. on Visualization and Computer Graphics},
  25(1):364--373, 2018.

\bibitem{iForestZhao}
X.~{Zhao}, Y.~{Wu}, D.~L. {Lee}, and W.~{Cui}.
\newblock {iForest}: Interpreting random forests via visual analytics.
\newblock {\em IEEE Trans. on Visualization and Computer Graphics},
  25(1):407--416, Jan 2019.

\bibitem{zheng2017machine}
T.~Zheng, W.~Xie, L.~Xu, X.~He, Y.~Zhang, M.~You, G.~Yang, and Y.~Chen.
\newblock A machine learning-based framework to identify type 2 diabetes
  through electronic health records.
\newblock {\em Int. J. Medical Informatics}, 97:120--127, 2017.

\end{thebibliography}
